\documentclass[twocolumn,prb,superscriptaddress]{revtex4}

\usepackage{picinpar}
\usepackage{array}
\usepackage{subfigure}
\usepackage{amsmath}
\usepackage{multirow}
\usepackage[dvips]{graphicx}
\usepackage{amssymb}

\begin{document}

\title{Donor and acceptor levels of organic photovoltaic compounds from first principles}

\author{Ismaila Dabo}
\email[corresponding author: ]{daboi@cermics.enpc.fr}
\affiliation{Universit\'e Paris-Est, CERMICS, Project-team INRIA Micmac, 
6 \& 8 avenue Blaise Pascal, 77455 Marne-la-Vall\'ee, France}
\author{Andrea Ferretti}
\affiliation{Centro S3, CNR--Istituto Nanoscienze, I-41125 Modena, Italy.}
\author{Cheol-Hwan Park}
\affiliation{Theory and Simulations of Materials, \'Ecole Polytechnique F\'ed\'erale de Lausanne, Station 12, 1015 Lausanne, Switzerland}
\affiliation{Department of Physics and Astronomy, Seoul National University, Seoul, Korea}
\author{Nicolas Poilvert}
\affiliation{Rowland Institute at Harvard, Cambridge, USA}
\author{Yanli Li}
\affiliation{Department of Physics, Xiamen University, Xiamen 361005, Republic of China}
\author{Matteo Cococcioni}
\affiliation{Department of Chemical Engineering and Materials Science,
University of Minnesota, Minneapolis, MN, USA}
\author{Nicola Marzari}
\affiliation{Theory and Simulations of Materials, \'Ecole Polytechnique F\'ed\'erale de Lausanne, Station 12, 1015 Lausanne, Switzerland}


\begin{abstract}
Accurate and efficient approaches to predict the optical properties of organic semiconducting compounds could accelerate the search for efficient organic photovoltaic materials. Nevertheless, predicting the optical properties of organic semiconductors has been plagued by the inaccuracy or computational cost of conventional first-principles calculations. In this work, we demonstrate that orbital-dependent density-functional theory based upon Koopmans' condition [Phys. Rev. B 82, 115121 (2010)] is apt at describing donor and acceptor levels for a wide variety of organic molecules, clusters, and oligomers within a few tenths of an electron-volt relative to experiment, which is comparable to the predictive performance of many-body perturbation theory methods at a fraction of the computational cost.
\end{abstract}

\maketitle

\section{Introduction}

\begin{figure}[ht!]
\includegraphics[width=8.5cm]{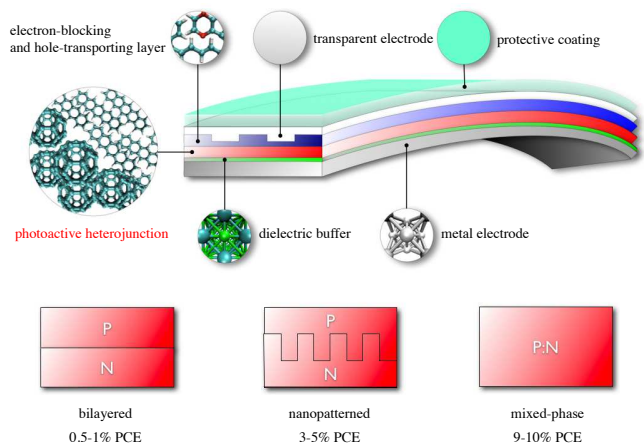}
\caption{Schematics of an organic solar cell showing a bilayered donor-acceptor heterojunction (top). Bilayered, nanopatterned, and mixed-phase heterojunctions with typical power conversion efficiencies (bottom).
\label{OSC}}
\end{figure}

Despite the positive attributes of organic materials for next-generation photovoltaics, the effective deployment of organic photovoltaic (OPV) cells poses fundamental problems at the molecular scale. Besides enhancing the durability of organic compounds, the central requisite to the mass-market viability of OPV modules is to raise their power conversion efficiency (PCE) beyond 10-15\%. \cite{HachmannOlivares-Amaya2011} Comparatively, the efficiency of current OPV architectures (Fig.~\ref{OSC}) barely exceeds 9-10\% PCE under standard test conditions. \cite{DouYou2012}  Nevertheless, the margin for improvement is vast due to the exceptional chemical versatility of organic materials, \cite{ScharberMuhlbacher2006} and accurate approaches to predict the charge-transfer and optical properties of new compounds could allow for extensive screening of promising semiconducting organics. \cite{SokolovAtahan-Evrenk2011}

However, conventional density-functional theory approximations in their static (DFT) and time-dependent (TDDFT) formulations \cite{HohenbergKohn1964, KohnSham1965, RungeGross1984} suffer from severe limitations in predicting the electronic structure of semiconducting materials, \cite{CohenMori-Sanchez2008} precluding the quantitative and even qualitative elucidation of relevant photovoltaic mechanisms. Notably, conventional DFT and TDDFT approximations tend to systematically destabilize occupied states and overstabilize unoccupied states in organic and organometallic complexes, leading in particular to the incorrect description of the electronic properties (charged excitations) and optical properties (neutral excitations) of donor-acceptor dyads, extended polymer molecules, and heavy-metal sensitizing dyes. \cite{DreuwWeisman2003, FaassenBoeij2002, BlaseAttaccalite2011a, HimmetogluMarchenko2012} To overcome DFT limitations, one privileged route has been to resort to many-body approaches, namely, to many-body perturbation theory approximations such as GW. \cite{Hedin1965,OnidaReining2002} The GW method, while accurate in predicting electronic spectra, is much more expensive than DFT although considerable progress has been achieved in reducing the cost of GW calculations. \cite{UmariStenuit2009,UmariStenuit2010, BlaseAttaccalite2011,BlaseAttaccalite2011a,FaberAttaccalite2011,QianUmari2011}

In parallel to many-body perturbation theory, less expensive orbital-dependent density-functional theories (OD-DFTs) represent promising alternatives. \cite{KummelKronik2008} At present, the most widely used OD-DFT methods are hybrid density-functional theory  approximations. \cite{Becke1993} Hybrid functionals in their linear-admixture or range-separated forms have been shown to improve upon conventional DFT in predicting electronic properties. \cite{KummelKronik2008, VarsanoAndrea-Marini2008, BroqvistAlkauskas2009, SteinEisenberg2010} Alternatively, self-interaction corrections to density-functional theory  approximations \cite{PerdewZunger1981} represent a second category of OD-DFT methods that are now rapidly growing in recognition due to their improved accuracy and lower computational cost. In self-interaction-corrected approaches, the total energy of the system is expressed explicitly in terms of individual orbital densities to rectify nonphysical errors inherent in orbital-independent DFT. Successful applications of advanced self-interaction corrections have appeared with an accuracy potentially approaching that of many-body methods in predicting electronic levels and electrical responses. \cite{BaerNeuhauser2005, KorzdorferMundt2008, DaboFerretti2010, FerrettiDabo2012}

In this work, we highlight the predictive accuracy of the recently developed self-interaction correction based upon Koopmans' theorem (Koopmans-compliant OD-DFT) \cite{DaboFerretti2010, FerrettiDabo2012} in capturing the electronic levels of OPV compounds. Electronic levels represent reliable PCE indicators \cite{ScharberMuhlbacher2006} and have successfully served as fundamental inputs for first-principles combinatorial screening of organic semiconductors. \cite{HachmannOlivares-Amaya2011, SokolovAtahan-Evrenk2011} We demonstrate that Koopmans-compliant OD-DFT is apt at describing donor levels within 0.1-0.4 eV and acceptor levels within 0.2-0.6 eV relative to experiment, which is comparable to the precision of the GW method. In its simplest formulation, the method can be trivially implemented in conventional electronic-structure codes with an algorithmic cost potentially lower than that of hybrid-DFT functionals as it does not require evaluating exchange terms. Furthermore, the method holds promise for accurate time-dependent extensions based upon the improved description of the underlying electronic spectrum.

This work is organized as follows. First, we underscore the significance of Koopmans' theorem in understanding the charge-transfer behavior of donor-acceptor complexes. Second, we outline existing electronic-structure approaches that aim at enforcing Koopmans' theorem. Third, we present the Koopmans-compliant OD-DFT method to correct the nonphysical tendency of conventional DFT approximations to destabilize occupied states and overstabilize unoccupied states. Last, we demonstrate the efficiency and accuracy of the method for families of organic compounds of interest to photovoltaics.

\section{Method} 

\label{Methodology}

\subsection{Koopmans' theorem}

\begin{figure}[ht!]
\includegraphics[width=7.5cm]{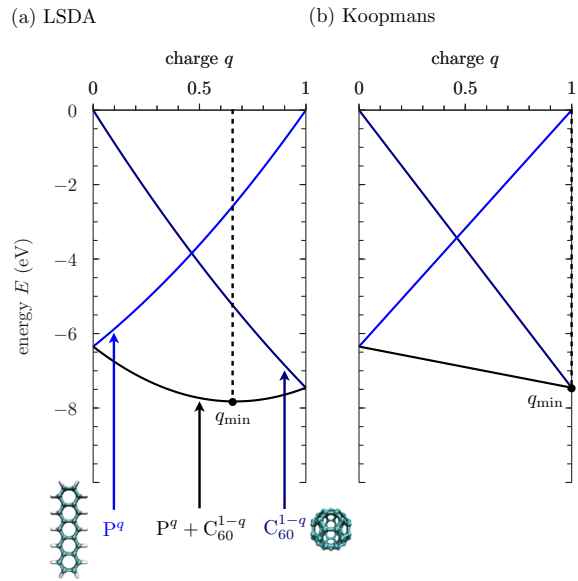}
\caption{Energy of a positively charged pentacene-fullerene dyad in the infinite separation limit (black) and energies of the isolated fractionally charged pentacene P$^q$ (light blue) and fullerene C$^{1-q}_{60}$ molecules (dark blue) as a function of the transferred charge $q$ within the LSDA and Koopmans-compliant descriptions. (Energies are given relative to the fully ionized individual molecules.)
\label{InfiniteSeparation}}
\end{figure}

Koopmans-compliant functionals aim at imposing Koopmans' theorem in DFT approximations. Koopmans' theorem enables one to equate orbital energies (that is, the expectation values of the effective Hamiltonian) with total energy differences, which correspond to withdrawing an electron from a stationary electron state.

 The relevance of Koopmans' theorem on the accuracy of effective-potential theories in predicting electronic spectra has been recognized in several theoretical and computational studies. \cite{HachmannOlivares-Amaya2011, SteinEisenberg2010, BaerLivshits2010, FerreiraMarques2011, AlkauskasPasquarello2011, JarlborgBarbiellini2011, CohenMori-Sanchez2012, KronikStein2012} In addition, Koopmans' theorem is the central condition to correctly describe electron transfer within donor-acceptor pairs. To illustrate this fact, we consider a positively charged pentacene-fullerene dyad in the limit of infinite intermolecular separation where electronic interactions between the two molecules can be neglected. 
 
Figure \ref{InfiniteSeparation} depicts the energies of the isolated pentacene oligomer and fullerene cluster. The total energy of the infinitely separated molecular pair as a function of the transferred charge $q$ is also reported. Within the local spin-density approximation (LSDA) [Fig.~\ref{InfiniteSeparation}(a)], the three curves exhibit a marked nonlinear dependence on $q$, reflecting the fact that LSDA violates Koopmans' theorem. In other words, the derivative of the energy $\textrm{d}E(\textrm{X}^q)/\textrm{d}q$ of an individual fractionally charged cation X$^q$, which corresponds to the  opposite energy of its highest occupied orbital, departs significantly from the finite energy difference $E(\textrm{X}^+)-E(\textrm{X})$, which corresponds to the ionization potential of the neutral molecule X. More precisely, the dependence of the energies is found to be convex, causing the dyad to be most stable at a fractional charge $q_\textrm{min}$ close to $\frac 23$. This observation is in qualitative contradiction to the expectation that the charge should fully localize onto the strongest electron acceptor (namely, fullerene). In contrast to this nonphysical behavior, Koopmans-compliant energy curves are linear [Fig.~\ref{InfiniteSeparation}(b)] as a result of Koopmans' condition $\textrm{d}E(\textrm{X}^q)/\textrm{d}q=E(\textrm{X}^+)-E(\textrm{X})$, leading to the expected stabilization of the fully transferred charge ($q_\textrm{min}=1$).

Similar observations would be made for other donor-acceptor complexes. Moreover, at finite separation where electronic coupling between the two molecules must be taken into account, LSDA would also significantly underestimate the transferred charge. These results highlight the central importance of imposing Koopmans' theorem on the individual subsystems to correctly describe charge transfer in donor-acceptor pairs.

\subsection{Existing methods}

\label{ExistingMethods}

At present, there exist different DFT-based methods that aim at imposing Koopmans' theorem. \cite{Dabo2008, SalznerBaer2009, LanyZunger2009, SteinEisenberg2010, BaerLivshits2010, LanyZunger2010, Refaely-AbramsonBaer2011} The self-interaction correction of Perdew and Zunger \cite{PerdewZunger1981} can be regarded as one of the earliest approaches to restore the physical interpretation of orbital energies as ionization energies. In this approach, the Kohn-Sham density functional $E^{\rm KS}$ is augmented with corrections that remove the self-interaction of the electronic wave functions in the one-electron limit. In explicit terms, the Perdew-Zunger (PZ) correction to the Kohn-Sham DFT functional reads $E^{\rm PZ}=E^{\rm KS}-\sum_{i\sigma} E^{\rm KS}_{\rm Hxc}[\rho_{i\sigma}]$, where $\rho_{i\sigma}$ stands for the density of the one-electron state $\psi_{i\sigma}$ and $E^{\rm KS}_{\rm Hxc}$ denotes the Hartree plus exchange-correlation component of the Kohn-Sham functional. The PZ approach is more expensive than local and semilocal DFT approximations but appreciably less costly than hybrid-DFT approximations. It rectifies the tendency of local and semilocal DFT to destabilize occupied orbitals in atoms but overcorrects the error for molecules, leading to an overestimation of ionization energies of up to 1-2 eV. Also, the PZ correction leaves the energies of empty levels unchanged, meaning that acceptor levels remain largely underestimated (that is, excessively negative). Furthermore, the lack of balance in the PZ correction of occupied and unoccupied orbitals leads to the underestimation of the total energies of molecules and the underestimation of interatomic distances. 

A solution to the lack of balance in the PZ correction has been proposed by Baer et al. in the form of a range-separated hybrid functional [the Baer-Neuhauser-Livshits (BNL) functional] in which long-range interactions are treated at the explicit-exchange level, whereas short-range interactions are described at the semilocal density-functional level. \cite{BaerNeuhauser2005,Livshitsbaer2007} In order to impose Koopmans' theorem within BNL, the original empirical parameter $\gamma$ that determines the range of the electrostatic separation ($1/r={\rm erf}(\gamma r)/r + {\rm erfc}(\gamma r)/r$) is optimally tuned to impose the agreement between orbital energies and ionization energies for the frontier orbitals of the system, \cite{SalznerBaer2009,BaerLivshits2010} leading to considerable improvement in predicting electronic and optical spectra. \cite{SteinEisenberg2010, Refaely-AbramsonBaer2011} In terms of computational cost, BNL calculations are as expensive as hybrid-DFT calculations.

A recently proposed approach that targets minimal computational cost is the method of Lany and Zunger, \cite{LanyZunger2009, LanyZunger2010} which consists of rectifying deviations from Koopmans' theorem through on-site projection terms that shift the electronic levels in order to impose the correspondence between orbital energies and ionization energies. Because it only relies on the calculation of on-site occupations, the Lany-Zunger method does not cause any noticeable increase in computational cost relative to the underlying DFT approximation. However, it requires to preselect atomic orbitals to construct the on-site projectors, making its application less systematic than the PZ correction and optimally tuned BNL method.

\subsection{Present method}

The Koopmans-compliant method that we have introduced in Ref.~\onlinecite{DaboFerretti2010} provides a promising alternative to existing approaches. The method is systematic and allies conceptual simplicity, computational performance, and predictive accuracy, thereby permitting the precise and efficient electronic-structure description of a wide variety of compounds. 

The motivation for introducing the Koopmans-compliant method originates from the practical observation that determining ionization potentials (IPs) of molecules as differences of ground-state energies $I_N=E_{N-1}-E_N$ (where $I_N$ and $E_N$ denote the IP and total energy of the system of $N$ electrons) yields predictions in very good agreement with experiment within conventional DFT approximations, namely, the LSDA approach and semilocal generalized-gradient approximations (GGAs) --- in the literature, the procedure that consists in evaluating ionization energies from differences of DFT total energies is commonly known as the $\Delta$SCF method. \cite{KowalczykYost2011} For electron affinities (EAs), $\Delta$SCF predictions $A_N=E_N-E_{N+1}$ are also found to be in agreement with experiment with the central caveat that $\Delta$SCF predictions for EAs are only possible insofar as the system can accommodate the added electron within the approximation used; this condition on the stability of anionic states is necessary to calculate the ground-state energy of negatively charged systems. In practical terms, failure to stabilize anions is frequent within approximate DFT \cite{KimSim2011} and arises from spurious self-interaction whereby an electron can interact with itself through the nonphysical contribution from its own charge density to the effective Kohn-Sham Hamiltonian. \cite{PerdewZunger1981}

Therefore, in order to obtain reliable predictions for orbital levels (especially, for frontier donor and acceptor levels that determine the charge-transfer properties of the system), it is crucial to satisfy the following conditions: (a) the correspondence between the effective orbital levels and $\Delta$SCF total-energy differences must be imposed; (b) the established accuracy of $\Delta$SCF energies must be preserved; and (c) the calculations must circumvent the inability of approximate functionals to stabilize negatively charged systems. Fulfilling these requirements is the object of the non-Koopmans OD-DFT correction that is described below. 

\begin{figure}[ht!]
\includegraphics[width=8.5cm]{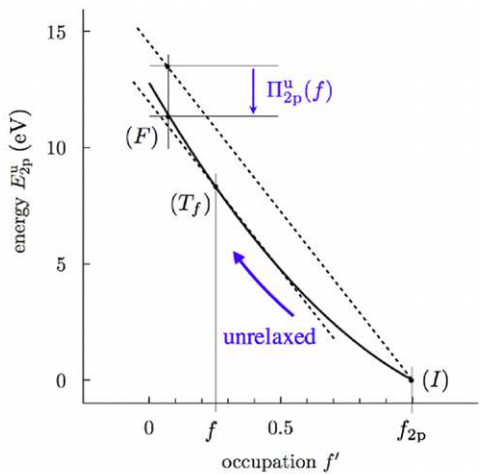}
\caption{Diagram of the non-Koopmans energy $\Pi^{\rm u}_{\rm 2p}$, which quantifies  departure from linearity for the adiabatic ionization curve of the 2p state of carbon within LSDA.
\label{CarbonAdiabatic}}
\end{figure}

In presenting the non-Koopmans correction, it is important to distinguish between electronically adiabatic (i.e., unrelaxed) ionization processes, whereby orbitals are kept frozen upon ionizing the system, and diabatic (i.e., relaxed) ionization processes, whereby orbitals are allowed to rearrange self-consistently. For adiabatic ionization, the correspondence between orbital energies and total energies is referred to as the restricted Koopmans' theorem, whereas for diabatic ionization, it is termed the generalized Koopmans' theorem.\cite{SzaboOstlund1989} In what follows, we consider the adiabatic case before addressing the diabatic case.

Bearing in mind this distinction, the starting point of the method consists in defining an adiabatic measure of the deviation from Koopmans' theorem in terms of an orbital-dependent energy functional. We thus consider for the moment an unrelaxed process that corresponds to depleting the stationary state $\psi_{i\sigma}$. For this process, the energy measure that monitors deviations from the restricted Koopmans' theorem can be written as the non-Koopmans energy
\begin{equation}
\Pi^{\rm u}_{i\sigma}(f)=\int_0^{f_{i\sigma}} {\rm d}f' [\epsilon^{\rm u}_{i\sigma}(f) - \epsilon^{\rm u}_{i\sigma}(f')],
\label{NKEa}
\end{equation}
first introduced by Perdew and Zunger in discussing the nonlinear behavior of the DFT total energy upon ionization. \cite{PerdewZunger1981} In Eq.~\eqref{NKEa}, the superscript notation  $\{\cdot\}^{\rm u}$ indicates that the orbitals are kept unrelaxed along the ionization path, $f_{i\sigma}$ denotes the initial occupation of the ionized orbital, and $\epsilon^{\rm u}_{i\sigma}$ stands for the expectation value of the effective Kohn-Sham Hamiltonian for the ionized state $\psi_{i\sigma}$ along the unrelaxed curve. In other words, $\epsilon^{\rm u}_{i\sigma}$ equals the derivative of the unrelaxed total energy $E^{\rm u}_{i\sigma}$ as a function of the occupation of the ionized state (the restricted Janak theorem). Alternatively, the unrelaxed non-Koopmans energy can be written in explicit terms as
\begin{eqnarray}
\Pi^{\rm u}_{i\sigma}(f)&=&f_{i\sigma}(2f-f_{i\sigma})E_{\rm H}[n_{i\sigma}]
+E_{\rm xc}[\rho-\rho_{i\sigma}] - E_{\rm xc}[\rho] \nonumber \\
&+& \int {\rm d}{\bf r} v_{{\rm xc},\sigma}\textstyle\left({\bf r};\left[\rho+\left(f - f_{i\sigma}\right)n_{i\sigma}\right]\right)\rho_{i\sigma}({\bf r}),
\label{NKEb}
\end{eqnarray}
where $\rho_{i\sigma}({\bf r})$ stands for the orbital density, $n_{i\sigma}({\bf r})=\rho_{i\sigma}({\bf r})/\int {\rm d}{\bf r}' \rho_{i\sigma}({\bf r}')$ denotes the normalized density, $E_{\rm xc}$ and $E_{\rm H}$ represent the exchange-correlation and Hartree energy functionals, and $v_{{\rm xc},\sigma}$ is the exchange-correlation potential.

The pictorial interpretation of the non-Koopmans energy is simple. Graphically, the unrelaxed non-Koopmans energy $\Pi^{\rm u}_{i\sigma}(f)$ corresponds to the error made in approximating the ionization curve connecting the initial state $(I)$ to the final state $(F)$ by a straight line whose slope is the derivative of the total energy at the transition point $(T_f)$ (Fig.~\ref{CarbonAdiabatic}). In fact, if the adiabatic Koopmans' theorem were satisfied, that is, if the energy $\epsilon^{\rm u}_{i\sigma}(f)$ were constant and equal to the $\Delta$SCF energy difference $E^{\rm u}_{i\sigma}(f_{i \sigma})-E^{\rm u}_{i\sigma}(0)$, the ionization curve would be exactly linear and the non-Koopmans deviation $\Pi^{\rm u}_{i\sigma}(f)$ would vanish for any occupation $f$ between 0 and $f_{i\sigma}$.

With the non-Koopmans measure in hand, it becomes possible to impose Koopmans' theorem, as shown in Ref.~\onlinecite{DaboFerretti2010}. Here, for the purpose of outlining the method, the non-Koopmans-corrected (NKC) functional can be described simply as a sum of weighted orbital constraints that are added to the Kohn-Sham total energy functional to penalize deviations from Koopmans' theorem:
\begin{equation}
E^{\rm NKC} = E^{\rm KS}+{\sum_{i\sigma}} \alpha^{\rm NK}_{i\sigma} \Pi^{\rm u, KS}_{i\sigma}(f^{\rm NK}_{i\sigma}),
\label{NKC}
\end{equation}
In Eq.~\eqref{NKC}, the terms $\Pi^{\rm u, KS}_{i\sigma}$ are the adiabatic non-Koopmans deviations that correspond to the chosen Kohn-Sham approximation, the coefficients $\alpha_{i\sigma}^{\rm NK}$ represent the weights of the non-Koopmans penalties, which must be calculated self-consistently to equate orbital energies with $\Delta$SCF energy differences [condition (a)] and the coefficients $f^{\rm NK}_{i\sigma}$ stand for the reference occupations, which must be fixed to preserve the accuracy of $\Delta$SCF energies [condition (b)], as explained further below.

At this point, it is important to note that Kohn-Sham density-functional theory is a ground-state theory, which is originally not intended to predict excited-state energies. As a consequence, only the non-Koopmans energies of the highest occupied and lowest unoccupied molecular orbitals can be rigorously defined within DFT since they correspond to ionization processes that take the $N$-electron ground state into the ground states with $N-1$ and $N+1$ electrons, respectively. In contrast, the ionization of other electron orbitals ends up into non-Aufbau states whose non-Koopmans energies are in principle not defined within DFT. 

As a consequence, the OD-DFT penalty sum that appears in Eq.~\eqref{NKC} should be rigorously restricted to the frontier highest occupied and lowest unoccupied states. Despite this recognized limitation, it is realized in practical computations that $\Delta$SCF based upon local and semilocal DFT functionals predict accurate excitation energies provided that orbitals do not delocalize nonphysically. \cite{KowalczykYost2011,HimmetogluMarchenko2012} These computational observations indicate that generalizing non-Koopmans penalties to the full orbital spectrum in Eq.~\eqref{NKC} should yield a Koopmans-compliant OD-DFT method able to predict electronic structures with good accuracy. In practice, it is found that this generalization provides spectral predictions in very good agreement with experimental data. \cite{DaboFerretti2010, FerrettiDabo2012}

\begin{figure}[ht!] 
\flushleft (a) unrelaxed \\
\center \includegraphics[width=5cm]{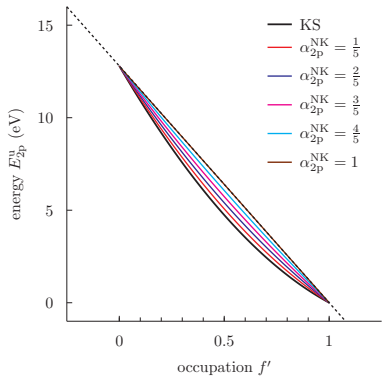} \\
\flushleft (b) relaxed \\
\center \includegraphics[width=5cm]{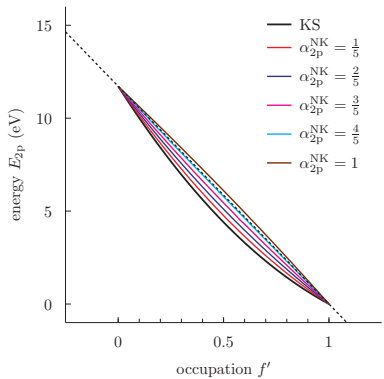}
\caption{Occupation dependence of the non-Koopmans-corrected total energy upon ionization of the 2p state of carbon as a function of the penalty coefficient $\alpha_{\rm 2p}^{\rm NK}$ in the (a) unrelaxed and (b) relaxed cases. The underlying Kohn-Sham approximation is LSDA. The reference occupation is of $f_{\rm 2p}^{\rm NK} = \frac 12$.
\label{CarbonAlpha}}
\end{figure}

To complete the presentation of the Koopmans-compliant functional, it remains to explain the determination of the penalty coefficients $\alpha_{i\sigma}^{\rm NK}$ and reference occupations $f^{\rm NK}_{i\sigma}$. The analytical determination of $\alpha_{i\sigma}^{\rm NK}$ has already been presented. \cite{DaboFerretti2010} In this analysis, it has been shown that, starting from convex-energy local and semilocal DFT functionals, there always exist a value of $\alpha_{i\sigma}^{\rm NK}$ lying between 0 and 1 that equalizes the slope of the ionization curve (i.e., the orbital energies) at the initial point $(I)$ and that at the final point $(F)$. Furthermore, the value of $\alpha_{i\sigma}^{\rm NK}$ that fulfills this condition can be calculated straightforwardly through the secant-method recursion:
\begin{equation}
\alpha_{n+1} = \alpha_n + \frac{(1-\alpha_n)\Delta^{\rm NK}_{i\sigma}(\alpha_n)}{\Delta^{\rm  NK}_{i\sigma}(\alpha_n)-\Delta^{\rm  NK}_{i\sigma}(1) },
\label{ALP}
\end{equation}
where $\Delta^{\rm NK}_{i\sigma}(\alpha)$ denotes the difference of orbital energies from $(I)$ to $(F)$ for the penalty coefficient $\alpha^{\rm NK}_{i\sigma}=\alpha$. At that precise value, the energy curve is closely linear and Koopmans' theorem is satisfied. 

To illustrate this fact, we depict the influence of $\alpha_{i\sigma}^{\rm NK}$ on the linearity of the energy curve upon ionization of the 2p state of carbon in Fig.~\ref{CarbonAlpha}. Considering first the adiabatic case [Fig.~\ref{CarbonAlpha}(a)], it is seen that the linearity of the ionization energy, that is, the fulfilment of the adiabatic Koopmans' theorem, is achieved for $\alpha_{\rm 2p}^{\rm NK}=1$. This result remains valid for any adiabatic process, as demonstrated by analyzing non-Koopmans residual errors. \cite{DaboFerretti2010} 

Instead, in the diabatic case, the value of $\alpha_{i\sigma}^{\rm NK}$ that satisfies the generalized Koopmans' theorem is found to be lower than 1. As a matter of fact, for relaxed ionization of the 2p state of carbon, it is seen in Fig.~\ref{CarbonAlpha}(b) that linearity is achieved for $\alpha_{\rm 2p}^{\rm NK}$ close to $\frac 45$. Precisely, the converged value of the penalty coefficient that is determined after 2 recursion steps [Eq.~\eqref{ALP}] is of 0.85. More generally, it has been shown that  $\alpha_{i\sigma}^{\rm NK}$ can be regarded as a relaxation coefficient that measures the stability of the final ionized state. Thus, for systems that consist of a lone electron around a closed shell (e.g., alkali-metal atoms), the outer-shell penalty coefficient equals 1. In contrast,  $\alpha_{i\sigma}^{\rm NK}$ becomes lower than 1 upon withdrawing an electron from a filled or partly filled shell (cf.~Fig.~9 in Ref.~\onlinecite{DaboFerretti2010}).

\begin{figure}[ht!]
\includegraphics[width=7cm]{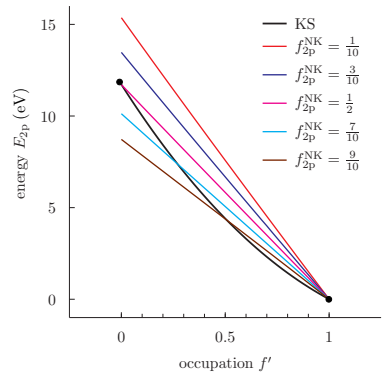}
\caption{Occupation dependence of the non-Koopmans-corrected total energy upon relaxed ionization of the 2p state of carbon as a function of the reference occupation $f_{\rm 2p}^{\rm NK}$. The underlying Kohn-Sham approximation is LSDA. The penalty coefficient is set to be $\alpha_{\rm 2p}^{\rm NK}=0.85$, as calculated recursively from Eq.~\eqref{ALP}.
\label{CarbonReferenceOccupation}}
\end{figure}

After explaining the calculation of the penalty coefficients, we now turn our attention to the determination of the reference occupation. Based upon Slater's theorem and the sum rule satisfied by the exchange-correlation hole, it can be shown that $f_{i\sigma}^{\rm NK}$ should be set to $\frac 12$ in order to preserve the accuracy of the underlying Kohn-Sham functional in predicting total-energy differences. \cite{DaboFerretti2010} This fact is illustrated in Fig.~\ref{CarbonReferenceOccupation} where it can be seen that the Koopmans-compliant energy difference matches the Kohn-Sham $\Delta$SCF energy difference at half reference occupation for diabatic ionization of the 2p state of carbon. Therefore, at the reference occupation $f^{\rm NK}_{\rm 2p}=0.5$ and with the relaxation parameter $\alpha_{\rm 2p}^{\rm NK}=0.85$ that has been determined from Eq.~\eqref{ALP}, the Koopmans-compliant functional yields an orbital energy $\epsilon_{\rm 2p}^{\rm NK}=11.83$ eV on a par with the accuracy of the $\Delta$SCF prediction (11.72 eV) and experimental IP (11.26 eV).

These results demonstrate the possibility of rectifying the nonlinear behavior of the ionization curve and restoring Koopmans' theorem [condition (a)] while preserving the accuracy of total-energy predictions [condition (b)] within Koopmans-compliant OD-DFT. In addition, Koopmans compliance presents the central advantage of avoiding to treat negatively charge molecular states [condition (c)] for both the evaluation of non-Koopmans penalties and the calculation of orbital levels, thereby allowing us to determine accurate acceptor levels straightforwardly, as shown in Sec.~\ref{BenchmarkPredictionsAcceptor}.

Further details on the implementation of one Koopmans-compliant method (the $\alpha$NKC$_0$ computational approach) are presented in the Appendix.

\section{Results}

\label{BenchmarkPredictions}

We now assess the predictive ability of Koopmans-compliant functionals in describing the donor-acceptor properties of a range of organic molecules relevant to OPV molecular junctions, first focusing on donor levels in Sec.~\ref{BenchmarkPredictionsDonor} and then examining the problem of acceptor levels in Sec.~\ref{BenchmarkPredictionsAcceptor}.

\subsection{Donor levels}

\label{BenchmarkPredictionsDonor}

\begin{figure}[ht!]
\includegraphics[width=8.5cm]{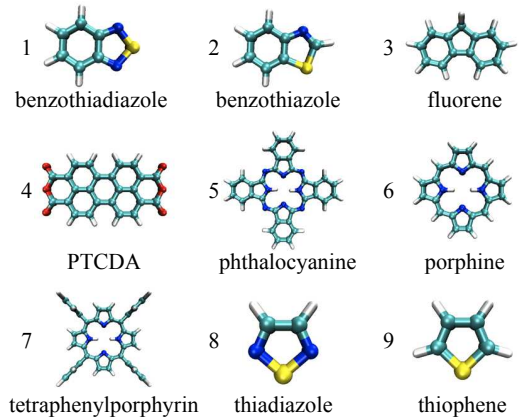}
\caption{Representative sample of organic molecules (Ref.~\onlinecite{BlaseAttaccalite2011}) for benchmarking the performance of electronic-structure methods.
\label{OrganicMolecules}}
\end{figure}

\begin{figure}[ht!]
\includegraphics[width=8.5cm]{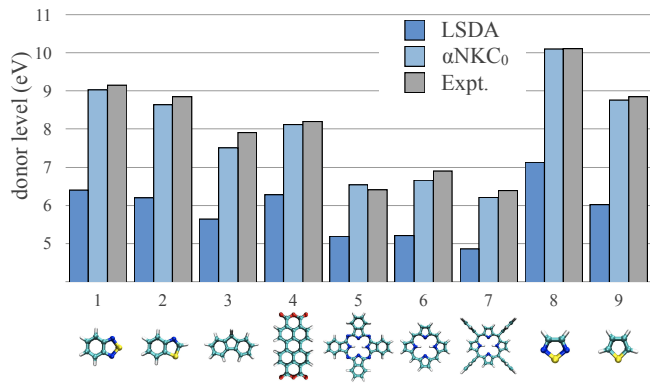}
\caption{LSDA and $\alpha$NKC$_0$ donor levels compared with experiment (cf.~references in Table \ref{OrganicMoleculesDonorComparison}) for the organic molecules listed in Fig.~\ref{OrganicMolecules}.
\label{OrganicMoleculesDonor}}
\end{figure}

\begin{table}
\caption{Absolute donor levels of benchmark organic molecules within LSDA, PZ, $\alpha$NKC$_0$, and GW compared with experimental vertical ionization energies (cf.~Ref.~\onlinecite{NIST2012} and references below).}
\label{OrganicMoleculesDonorComparison}
\begin{tabular*}{1.0\columnwidth}{@{\extracolsep{\fill}}lccccc}
\hline \hline \\
& \multicolumn{5}{c}{Donor levels (eV)} \\ 
\cline{2-6} \\
& LSDA & PZ & $\alpha$NKC$_0$ & GW\footnotemark[1] & Expt.\\ 
\hline
\\
benzothiadiazole & 6.40 &10.29& 9.03 & 8.56 & 9.15\footnotemark[2] \\ 
benzothiazole & 6.20 & 9.66 & 8.64 & 8.48 & 8.85\footnotemark[3] \\
fluorene & 5.64 & 8.69 & 7.51 & 7.64 & 7.91\footnotemark[4]\\
PTCDA & 6.32 & 9.66 & 8.12 & 7.68 & 8.2\footnotemark[5] \\
phthalocyanine & 5.21 & 8.31 & 6.54 & 6.10 & 6.41\footnotemark[6] \\
porphine & 5.24 & 8.34 & 6.66 & 6.70 & 6.9\footnotemark[7] \\
tetraphenylporphyrin & 4.89 & 8.10 & 6.21 & 6.20 & 6.39\footnotemark[8] \\
thiadiazole & 7.13 &11.73&10.10& 9.89 & 10.11\footnotemark[9]\\
thiophene & 6.02 &10.97& 8.76 & 8.63 & 8.85\footnotemark[10]\\
\\
MAD & 2.20  & 1.43 & 0.16 & 0.32 \\
RMS & 0.64 & 0.45 & 0.11 & 0.15 \\
\hline \hline
\end{tabular*}
\flushleft
\footnotemark[1]{Reference \onlinecite{BlaseAttaccalite2011}.} \\
\footnotemark[2]{Reference \onlinecite{JohnstoneMellon1973}.} \\
\footnotemark[3]{Reference \onlinecite{SalmonaFaure1975}.} \\
\footnotemark[4]{Reference \onlinecite{RuscicKovac1978}.} \\
\footnotemark[5]{Reference \onlinecite{DoriMenon2006}.} \\
\footnotemark[6]{Reference \onlinecite{Berkowitz1979}.} \\
\footnotemark[7]{Reference \onlinecite{DupuisRoberge1980}.} \\
\footnotemark[8]{Reference \onlinecite{KhandelwalRoebber1975}.} \\
\footnotemark[9]{Reference \onlinecite{PasinszkiKrebsz2010}.} \\
\footnotemark[10]{Reference \onlinecite{BajicHumski1985}.} \\
\end{table} 

\begin{figure}[ht!]
\includegraphics[width=8.5cm]{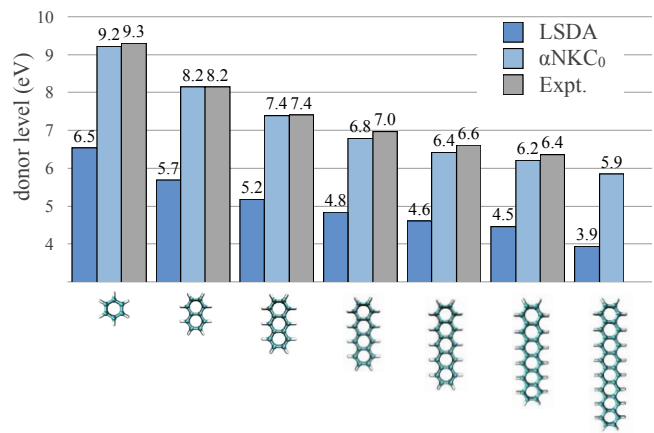}
\caption{LSDA and $\alpha$NKC$_0$ donor levels compared with experiment for benzene (Ref.~\onlinecite{PiancastelliKelly1987}) and oligoacenes (Ref. ~\onlinecite{BiermannSchmidt1980}).
\label{AcenesDonor}}
\end{figure}

\begin{figure}[ht!]
\includegraphics[width=8.5cm]{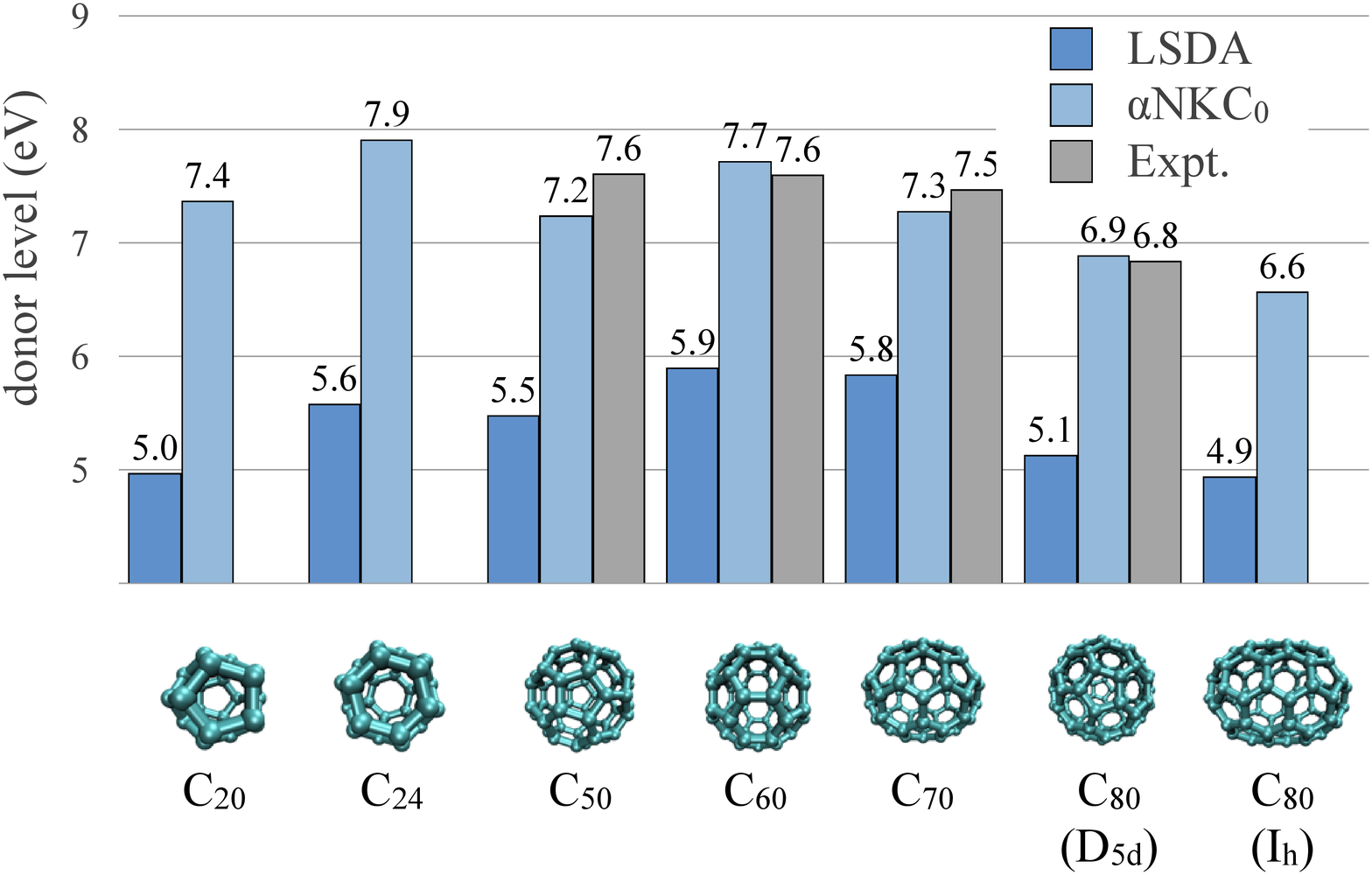}
\caption{LSDA and $\alpha$NKC$_0$ donor levels compared with experiment for fullerenes. Experimental vertical ionization potential are from Ref.~\onlinecite{ZimmermanEyler1991} for C$_{50}$, C$_{60}$, and C$_{80}$ (D$_{\rm 5d}$) and from Ref.~\onlinecite{LichtenbergerRempe1992} for C$_{70}$.
\label{FullerenesDonor}}
\end{figure}

The initial benchmark that is depicted in Fig.~\ref{OrganicMolecules} consists of a representative set of molecules, which has been studied in Ref.~\onlinecite{BlaseAttaccalite2011}. These molecules constitute elemental components for relevant photovoltaic organics and allow us to assess the accuracy of the self-interaction method in capturing the electronic structure of organic materials across a representative sample of chemical compositions and molecular sizes.

Koopmans-compliant calculations are carried out at the $\alpha$NKC$_0$ level. We employ conventional norm-conserving pseudopotentials to represent atomic cores with a cutoff energy of 60 Ry in expanding wave functions. All of our calculations take into account spin polarization. Employing auxiliary-function correction,\cite{LiDabo2011} a vacuum separation of 9 \AA\ is sufficient to achieve convergence of the electronic structure within 0.15 eV, even for the large PTCDA and phthalocyanine molecules. In all cases, less than two recursions of Eq.~\eqref{ALP} are needed in converging the penalty coefficient $\alpha$ of the highest occupied state of the molecule, and the same penalty weight is imposed on the full spectrum (as justified by the sensitivity analysis presented in the Appendix). Damped electronic dynamics is used in optimizing electronic degrees of freedom with the inner-loop procedure described in Ref.~\onlinecite{ParkFerretti2012} and with a convergence threshold of 10$^{-5}$ Ha on the electron-dynamics kinetic energy. Regarding the experimental literature, it is important to mention that all of the measurements that are reported here correspond to molecules isolated in the gas phase. Therefore, environmental effects arising from a surrounding solvent or from an embedding bulk need not be included in our calculations.  

LSDA and $\alpha$NKC$_0$ predictions for the frontier donor levels (i.e., the absolute energy of the highest occupied orbital) of benchmark molecules are reported and compared with experiment in Fig.~\ref{OrganicMoleculesDonor}. (Here and throughout the comparative assessment, we focus on LSDA results rather than on GGA predictions due to the fact that $\alpha$NKC$_0$ is applied to LSDA; this permits to directly resolve the effective influence of $\alpha$NKC$_0$ on the underlying functional. Note that GGA orbital energies would be very close to the LSDA results on the scale of the self-interaction error.) 

First, we observe a marked underestimation of absolute donor levels within LSDA. The underestimation is above 1 eV for porphyrin complexes and reaches 2.5-3 eV for strong aromatic donors. This incorrect trend is rectified by $\alpha$NKC$_0$ that reduces the error down to a few tenths of an eV. Furthermore, the accuracy of $\alpha$NKC$_0$ compares favorably with that of the original self-interaction correction (PZ)\cite{PerdewZunger1981} and GW approximation. Indeed, data reported in Table~\ref{OrganicMoleculesDonorComparison} reveal that PZ provides some improvement over LSDA, reducing the mean absolute error from 2.20 eV to 1.43 eV. Much better agreement with experiment is achieved within self-consistent GW\footnote{In these GW calculations, wave functions are kept frozen and self-consistency is performed on eigenenergies.} with a mean absolute error as low as 0.32 eV and a standard deviation of the error of 0.15 eV. Despite the very good performance of GW calculations,\footnote{It is beyond the scope of this work to enter into the delicate discussion of the accuracy of GW predictions. Notwithstanding recognized difficulties in converging GW calculations, the present comparisons provide a sufficiently precise assessment of the performance of first-principles methods to draw conclusions on their relative performance.}  the Koopmans-compliant $\alpha$NKC$_0$ method is found here to be more precise in predicting the highest donor levels with an error of 0.16 eV and a low standard deviation of 0.11 eV.

After highlighting the remarkable precision of $\alpha$NKC$_0$ across a variety of chemical compositions, we now assess its ability to capture the influence of molecular size on the electronic structure. For this complementary benchmark, we consider two important families of organic molecules, namely, acene oligomers and fullerene clusters.  We employ the same calculation parameters as those above --- with the exception of the plane-wave cutoff that can be reduced to 40 Ry without altering numerical convergence. 

For oligoacenes (Fig.~\ref{AcenesDonor}), we observe that LSDA is in error of 2.8 eV for the donor level of benzene. The error decreases gradually with the length of the chain but is still as large as 2 eV for hexacene. In terms of relative errors, the LSDA underestimation fluctuates in the narrow range 29-31\%. Applying $\alpha$NKC$_0$ restores the agreement between electronic-structure calculations and experiment, reducing the error to less than 0.2 eV regardless of the length of the chain. Similarly, for fullerenes (Fig.~\ref{FullerenesDonor}), $\alpha$NKC$_0$ brings about some considerable improvement in capturing the delicate influence of cluster size. Indeed, despite the fact that the LSDA error varies in a much wider range than for acene oligomers (22-28\%) and that the dependence of the highest occupied level as a function of the molecular size is clearly nonmonotonic, the $\alpha$NK$_0$ error never exceeds 0.2 eV compared to available experimental data, providing further confirmation of the predictive precision of the Koopmans-compliant method in describing subtle electronic-structure trends.

In addition, an important advantage of the $\alpha$NKC$_0$ method lies in its moderate computational cost. In fact, converging the full electronic spectrum of an extended PTCDA molecule with a large plane-wave cutoff of 60 Ry on a conventional 16-processor machine requires half of a CPU day, whereas the same calculation within HF and hybrid-DFT is 3 times more costly using comparable {\sc cp} implementations. Indeed, as explained previously, although conventional self-interaction corrections and hybrid-DFT both scale quartically [$O(N^4)$] with the size of the electronic system, the algorithmic prefactor of self-interaction methods is lower than that of hybrid-DFT since the former do not require to evaluate pairwise exchange contributions.

\subsection{Acceptor levels}

\label{BenchmarkPredictionsAcceptor}

\begin{table}
\caption{Absolute acceptor levels of benchmark organic molecules within LSDA, PZ, $\alpha$NKC$_0$, and GW compared with available experimental data.}
\label{OrganicMoleculesAcceptorComparison}
\begin{tabular*}{1.0\columnwidth}{@{\extracolsep{\fill}}lccccc}
\hline \hline \\
&  \multicolumn{5}{c}{Acceptor levels (eV)} \\ 
\cline{2-6} \\
& LSDA & PZ & $\alpha$NKC$_0$ & GW\footnotemark[1] & Expt. \\ 
\hline
\\
benzothiadiazole & 3.52 & 3.42 & 1.07  & 0.42 & ---  \\ 
benzothiazole & 2.35 & 2.20 & $<$0 & $<$0 & ---   \\
fluorene & 2.05 & 2.00 & $<$0 & $<$0 & ---  \\
PTCDA & 4.80 & 4.81 & 3.19  & 2.68 & ---  \\
phthalocyanine & 3.79 & 3.82 & 2.34  & 2.07 & ---  \\
porphine & 3.28 & 3.19 & 1.53  & 1.39 & ---  \\
tetraphenylporphyrin & 3.07 & 3.04 & 1.68  & 1.49  & 1.69(10)\footnotemark[2] \\
thiadiazole & 2.95 & 2.78 & $<$0 & $<$0 & --- \\
thiophene & 1.59 & 1.38 & 0.04  & $<$0 & --- \\
\\
\hline \hline 
\end{tabular*}
\flushleft
\footnotemark[1]{Reference \onlinecite{BlaseAttaccalite2011}.} \\
\footnotemark[2]{Reference \onlinecite{ChenPan1991}.}
\end{table} 

After validating the $\alpha$NKC$_0$ method for occupied donor levels, we now concentrate on the prediction of unoccupied acceptor levels. As already mentioned, acceptor levels represent a central difficulty for $\Delta$SCF predictions due to the inability of conventional DFT methods to properly stabilize negatively charged states. In what follows, we demonstrate that $\alpha$NKC$_0$ provides a solution to this important methodological limitation.

In calculating acceptor levels, we use the same computational parameters as those of donor-level calculations. In particular, the $\alpha$NK$_0$ penalty weights remain unchanged. First-principles predictions for benchmark organic compounds are reported in Table \ref{OrganicMoleculesAcceptorComparison}. Our results clearly confirm the poor performance of LSDA and PZ in predicting acceptor levels; both methods overstabilize acceptor states by several eVs relative to GW due to self-attraction. In contrast, $\alpha$NKC$_0$ and GW predictions are found to be in close agreement. In quantitative terms, the discrepancy between $\alpha$NKC$_0$ and GW is as low as 0.15 eV for porphine and is at most of 0.6-0.7 eV for benzothiadiazole. It should also be noted that our comparison reveals that $\alpha$NKC$_0$ acceptor levels are always more stable than their GW counterparts. Additionally, although experimental data for benchmark molecules are scarce, comparison with the measured vertical electron affinity of tetraphenylporphyrin suggests that the predictive accuracy of $\alpha$NKC$_0$ is comparable with the accuracy of GW predictions.

\begin{table}
\caption{Absolute acceptor levels of acene chains of increasing length within LSDA, $\alpha$NKC$_0$,and GW compared with experimental data.}
\label{AcenesAcceptor}
\begin{tabular*}{1.0\columnwidth}{@{\extracolsep{\fill}}lcccc}
\hline \hline \\
& \multicolumn{4}{c}{Acceptor levels (eV)} \\ 
\cline{2-5}\\
& LSDA & $\alpha$NKC$_0$ & GW\footnotemark[1] & Expt. \\ 
\hline
\\
benzene       & 1.38 & 0.03 & --- & --- \\ 
naphthalene & 2.27 & 0.49 & --- & --- \\
anthracene   & 2.85 & 1.16 & 0.29 & 0.53\footnotemark[2] \\
tetracene      & 3.22 & 1.60 & 0.93 & 1.07\footnotemark[3] \\
pentacene    & 3.48 & 1.98 & 1.36 & 1.39\footnotemark[3] \\
hexacene     & 3.68 & 2.18 & ---  & --- \\
octacene      & 4.27 & 2.48 & ---  & --- \\
\\
MAD & 2.19 & 0.59 & 0.13 \\
RMS & 0.09 & 0.04 & 0.09  \\
\hline \hline 
\end{tabular*}
\flushleft
\footnotemark[1]{Reference \onlinecite{BlaseAttaccalite2011}.} \\
\footnotemark[2]{Reference \onlinecite{SchiedtWeinkauf1997}.} \\
\footnotemark[3]{Reference \onlinecite{CrockerWang1993}.}
\end{table} 

\begin{table}
\caption{Absolute acceptor levels of fullerene clusters of increasing size within LSDA, $\alpha$NKC$_0$, QMC (quantum Monte-Carlo), GW$_0$ (nonself-consistent), scGW$_{\rm f}$ (self-consistent with vertex corrections) compared with experimental data.}
\label{FullerenesAcceptor}
\begin{tabular*}{1.0\columnwidth}{@{\extracolsep{\fill}}lcccccc}
\hline \hline \\
& \multicolumn{6}{c}{Acceptor levels (eV)} \\ 
\cline{2-7}\\
& LSDA & $\alpha$NKC$_0$ & QMC\footnotemark[1] & GW$_0$\footnotemark[1] & scGW$_{\rm f}$\footnotemark[1] & Expt. \\ 
\hline
\\
C$_{20}$ & 4.24 & 2.19 & 1.76(11) & 3.55 & 2.36 & 2.25\footnotemark[2] \\ 
C$_{24}$ & 5.06 & 2.88 & 2.57(11) & 4.19 & 2.88 & --- \\
C$_{50}$ & 5.20 & 3.49 & 3.52(14) & 4.75 & 3.73 & 3.10\footnotemark[3] \\
C$_{60}$ & 4.27 & 2.64 & 2.23(19) & 3.87 & 2.98 & 2.69\footnotemark[4]  \\
C$_{70}$ & 4.10 & 2.61 & 2.46(11) & 3.98 & 2.83 & 2.76\footnotemark[5] \\
C$_{80}$ (D$_{\rm 5d}$) & 5.00 & 3.91 & 3.25(10) & 4.62 & 3.88 & 3.70\footnotemark[3] \\
C$_{80}$ (I$_{\rm h}$)    & 4.56 & 2.99 & 3.90(11) & 5.17 & 4.38 & --- \\
\\
MAD & 1.66 & 0.18 & 0.42 & 1.25 & 0.26 \\
RMS & 0.32 & 0.12 & 0.06 & 0.26 & 0.21 \\
\hline \hline 
\end{tabular*}
\flushleft
\footnotemark[1]{Reference \onlinecite{TiagoKent2008}.} \\
\footnotemark[2]{Reference \onlinecite{PrinzbachWahl2006}.} \\
\footnotemark[3]{Reference \onlinecite{YangPettiette1987}.} \\
\footnotemark[4]{Reference \onlinecite{WangDing1999}.} \\
\footnotemark[5]{Reference \onlinecite{WangWoo2006}.}
\end{table} 

To gain further insight into the performance of first-principles methods in predicting acceptor levels, we concentrate on  the lowest unoccupied state for acenes (Table \ref{AcenesAcceptor}) and for fullerenes (Table \ref{FullerenesAcceptor}). For weak acene acceptors, the first notable feature is the propensity of LSDA to overestimate absolute acceptor levels. This expected nonphysical tendency is rectified to a large extent by $\alpha$NKC$_0$, which reduces the mean absolute error from 2.19 to 0.59 eV relative to vertical ionization experiments. Nevertheless, it is seen that GW performs significantly better than $\alpha$NKC$_0$, bringing the error down to 0.13 eV. In the same vein, we note that the naphthalene acceptor level is predicted to be stable within $\alpha$NKC$_0$ in contradiction to experiment. In fact, among all of our benchmark calculations, acenes represent the only instance where GW performs better than $\alpha$NKC$_0$, suggesting that GW should be preferred over $\alpha$NKC$_0$ in capturing shallow acceptor levels. 

Despite this fact, the scenario is entirely different for strong acceptors. Indeed, as shown in Table \ref{FullerenesAcceptor}, $\alpha$NKC$_0$ yields acceptor levels in close agreement with experimental data for fullerene clusters with a mean absolute deviation of 0.18 eV, which is significantly lower than that found for many-body quantum Monte-Carlo (0.42 eV), nonself-consistent many-body perturbation theory (1.25 eV), and self-consistent many-body perturbation theory (0.26 eV). These results suggest us to employ Koopmans-compliant $\alpha$NKC$_0$ to describe the electronic properties of the strong acceptor compounds relevant to photovoltaic applications. 

Overall, the systematic comparison of electronic-structure methods offers a clear validation of $\alpha$NKC$_0$ as an accurate and efficient method to describe the donor and acceptor levels of relevant OPV compounds. This remarkable performance allows us to access the electronic structure of novel organics and provides a reliable starting point to address the charge-transfer and optical properties of donor-acceptor photovoltaic junctions.

\section{Conclusion}

In order to overcome the predictive deficiency of conventional DFT methods in capturing the donor and acceptor levels of molecular complexes, we have presented a correction procedure that enables one to eliminate the nonphysical curvature of the total energy upon withdrawing (injecting) electrons from (into) the molecular system. When applied to LSDA, the procedure yields an orbital-dependent functional that fulfills the generalized Koopmans' theorem, thereby restoring or imposing the agreement between frontier orbital levels and $\Delta$SCF energy differences, and overcoming central limitations in applying $\Delta$SCF techniques to predict acceptor levels. Specifically, the $\alpha$NKC$_0$ formulation of the Koopmans-compliant method has been applied to OPV compounds spanning a wide range of chemical compositions and molecular sizes, thereby highlighting the conceptual simplicity and computational performance of the method with an accuracy comparable to that of many-body perturbation theory.

Although the study has essentially focused on determining frontier donor and acceptor levels, the non-Koopmans corrective procedure can be straightforwardly extended to compute the full electronic spectrum of molecular systems in a single first-principles calculation.\cite{FerrettiDabo2012} For future studies, we plan to apply the Koopmans-compliant method to other relevant families of photovoltaic compounds, including metallic porphyrins and phthalocyanines, and to generalize the approach to describe strongly interacting dyads and bulk organic semiconducting materials at both the electronic and excitonic levels.

\acknowledgments

The authors are indebted to Arash Mostofi, Xavier Blase, and Kieron Burke for valuable discussions and practical suggestions. ID acknowledges partial support from the French National Research Agency through Grant ANR 12-BS04-0001 PANELS (Photovoltaics from Ab-initio Novel Electronic-structure Simulations). AF acknowledges partial support from Italian MIUR through Grant FIRB-RBFR08FOAL\_001. MC acknowledges partial support from NSF EAR 0810272 and from the NSF CAREER award DMR 1151738.

\appendix

\section{Computational details}

\label{ComputationalDetails}

In this appendix, we present the computational implementation of the Koopmans-compliant method in the {\sc cp} (Car-Parrinello) code\cite{LaasonenPasquarello1993} of the {\sc quantum-espresso} open-source project. \cite{GiannozziBaroni2009} Specifically, we describe the calculation of contributions to the total energy and effective Hamiltonian that arise from the non-Koopmans correction, we explain algorithmic choices in optimizing the electronic structure, and we detail the evaluation of the penalty coefficients.

Before proceeding to the calculation of non-Koopmans contributions, it should be noted that {\sc cp} is a plane-wave code. Adopting a plane-wave representation here allows us to envision applications to relevant periodic molecular structures, such as semiconducting polymers and bulk materials. Despite the benefit of the plane-wave approach, the evaluation of electrostatic contributions to the non-Koopmans energies [Eq.~\eqref{NKEb}] in the context of plane waves requires to correct periodic-image errors that arise from the use of the supercell approximation in describing nonperiodic systems. To eliminate artificial periodic-image interactions, we employ auxiliary-function corrections that consist in removing singularities in the reciprocal-space summation of electrostatic terms through the addition of a regularizing auxiliary functions to the point-charge electrostatic kernel. \cite{LiDabo2011} The choice of this reciprocal-space scheme is motivated by the fact that it allows us to use fast-Fourier-transform (FFT) techniques, thereby minimizing the computational burden associated with the repeated calculations of the electrostatic energy and potential for each orbital of the system; without this correction the calculation would be prohibitively expensive.

The second difficulty in calculating non-Koopmans terms lies in the assessment of the second-order functional derivatives that appear in the expression of the variational contributions to the effective potential, namely, the $w^{\rm KS}_{{\rm ref},i\sigma}$ and  $w^{\rm KS}_{{\rm xd},i\sigma}$ terms defined in Ref.~\onlinecite{DaboFerretti2010}. To address the calculation of these terms, we employ the computational subroutines written by Dal Corso and de Gironcoli in the context of the DFPT (density-functional perturbation theory) calculation of phonon dispersions. \cite{Dal-CorsoGironcoli2000} These subroutines are based upon explicit expressions of second-order derivative contributions and also employ finite-difference techniques. Similar subroutines distributed as independent libraries (e.g., the {\sc libxc} module of the {\sc octopus} code)\cite{CastroAppel2006, Marques2003} can be conveniently ported to compute second-order terms. 

Overall, the calculation of non-Koopmans contributions to the total energy and effective Hamiltonian is already less expensive than that of exchange contributions in hybrid-DFT functionals.\cite{Becke1993} Nevertheless, in order to gain further computational performance, it is necessary to reduce the computational burden associated to the calculation of $w^{\rm KS}_{{\rm ref},i\sigma}$ and $w^{\rm KS}_{{\rm xd},i\sigma}$. To this end, a successful strategy consists in not updating the reference density $\rho_{i\sigma}^{\rm NK}({\bf r}) = \rho({\bf r})+(f_{i\sigma}^{\rm NK}-f_{i\sigma})n_{i\sigma}({\bf r})$ that appears in the expression of non-Koopmans energies at every iteration of the self-consistent electronic-structure optimization but only periodically for each given number of iterations, thereby avoiding to calculate $w^{\rm KS}_{{\rm ref},i\sigma}$ and  $w^{\rm KS}_{{\rm xd},i\sigma}$ at each self-consistent step. This computational approach that consists in neglecting the $w^{\rm KS}_{{\rm ref},i\sigma}$ and  $w^{\rm KS}_{{\rm xd},i\sigma}$ contributions to the self-consistent OD-DFT potential is referred to as the NKC$_0$ method. 

Remarkably, it is found that relaxing self-consistency via the NKC$_0$ method preserves and actually improves the accuracy of predicted electronic properties relative to NKC for difficult molecular complexes due to its increased localization strength, with the caveat that only orbital properties and not total energies can be evaluated using NKC$_0$.\cite{DaboFerretti2010} Variational alternatives to NKC$_0$, endowed with a comparable localization strength can be derived at the cost of increased sophistication. A preliminary discussion of variational non-Koopmans self-interaction corrections is presented in Ref.~\onlinecite{Poilvert2011}. In the present study, we focus on NKC$_0$ due to its simplicity and accuracy in describing individual organic donors and acceptors. Admittedly, more work would be necessary for the variational description of systems where strong electronic delocalization prevails.

Regarding computational implementation, NKC$_0$ is straightforward to program since it simply relies on substituting the DFT Kohn-Sham potential with the OD-DFT potential written below: 
\begin{eqnarray}
&  v_{\rm Hxc}^{\rm KS}[\rho]  \nonumber  \\
& \downarrow \label{NK0}\\
& (1-\alpha_{i\sigma}^{\rm NK})v^{\rm KS}_{\rm Hxc}[\rho] +\alpha^{\rm NK}_{i\sigma} v^{\rm KS}_{\rm Hxc}[\rho^{\rm NK}_{i\sigma}]   \nonumber
\end{eqnarray}
where $v^{\rm KS}_{\rm Hxc}$ denotes the sum of the Hartree and exchange-correlation potentials and $\rho_{i\sigma}^{\rm NK}$ has been defined above. Once the subroutines that calculate $v^{\rm KS}_{\rm Hxc}$ are identified, this modification can be performed with some minimal coding experience.

We now turn our attention to the optimization of the electronic structure. In this regard, we first note that only few of the conventional minimization algorithms applicable to DFT can be employed in the context of the OD-DFT corrections. In particular, conventional iterative diagonalization would represent a very poor choice since eigenfunctions of different Hamiltonians would have to be calculated for each of the orbitals in the system, thereby considerably increasing computational complexity. Instead, direct-minimization approaches, such as conjugate-gradient or damped-dynamics techniques, are more suited to OD-DFT calculations. Additionally, since OD-DFT self-interaction corrections lead to energy functional that are variant with respect to unitary rotation of the orbital manifold contrary to DFT, it is also necessary to consider unitary-rotation degrees of freedom in minimizing the energy. Several approaches have been proposed and can be used to this end. \cite{HeatonPederson1987, StengelSpaldin2008,KlupfelKlupfel2011,HofmannKlupfel2012, ParkFerretti2012}

The final hurdle in implementing the Koopmans-compliant method pertains to evaluating the penalty coefficients $\alpha_{i\sigma}^{\rm NK}$. 

\begin{figure}[ht!]
\includegraphics[width=7cm]{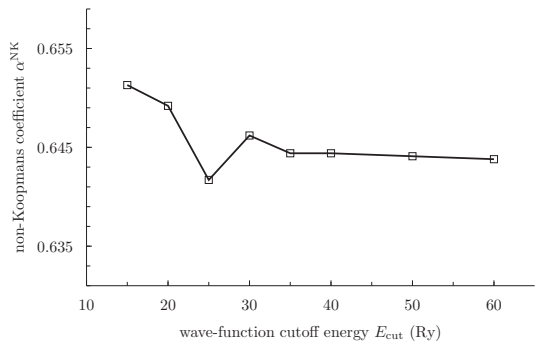}
\caption{Non-Koopmans penalty coefficient for the highest occupied molecular orbital [1a$_2$($\pi_3$)] of furan as a function of wave-function cutoff energy.
\label{FuranAlpha}}
\end{figure}

Here, one first source of computational savings springs from the fact that $\alpha_{i\sigma}^{\rm NK}$ is much less sensitive to the resolution of the plane-wave grid than orbital-level predictions. In quantitative terms, in the case of furan (Fig.~\ref{FuranAlpha}), reducing the cutoff from 60 Ry to 15 Ry causes the relaxation coefficient to vary by 3\%, whereas such a low plane-wave cutoff in expanding wave functions would be intolerably low for the description of orbital levels. Therefore, since $\alpha_{i\sigma}^{\rm NK}$ can be evaluated on coarse plane-wave grids,  the preliminary determination of the weights of the non-Koopmans penalty $\alpha_{i\sigma}^{\rm NK}$ represents a marginal fraction of the computational cost associated with the calculation of orbital levels.

\begin{figure}[ht!]
\includegraphics[width=7cm]{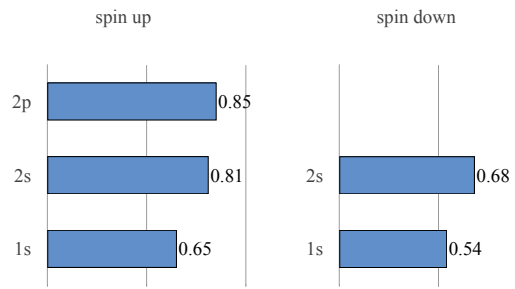}
\caption{Penalty coefficients $\alpha_{i\sigma}^{\rm NK}$ that quantify orbital relaxation upon ionization for the orbitals of carbon in the spin-up and spin-down channels. 
\label{LowLyingStates}}
\end{figure}

The second important source of cost reduction in practical calculations is the fact that the relaxation coefficients $\alpha_{i\sigma}^{\rm NK}$ varies in a limited range, typically, between 0.5 and 1. To illustrate this fact, we report the value of the coefficients $\alpha_{i\sigma}^{\rm NK}$ for the electronic states of carbon in Fig.~\ref{LowLyingStates}. On the basis of these calculations, we can assess the sensitivity of electronic-level predictions as a function of the $\alpha_{i\sigma}^{\rm NK}$'s. Indeed, LSDA is typically in error of $\Delta=40\%$ in predicting the electronic levels of atoms and molecules. As a consequence, a simple sensitivity estimate reveals that an error of $\alpha - \alpha_{i\sigma}^{\rm NK}$ in the determination of the self-consistent penalty coefficient translates into an error of $\frac{\alpha-\alpha_{i\sigma}^{\rm NK}}{\alpha_{i\sigma}^{\rm NK}} \Delta$ in the electronic level relative to the exact Koopmans-compliant estimate (that is, relative to the $\Delta$SCF energy difference). Thus, in the case of carbon, making the approximation of equating all of the penalty coefficients to that of the highest occupied 2p state (i.e., $\alpha=0.85$), we obtain errors as low as 2\% for the spin-up 2s state and lower than $14$\% for the spin-down 1s state, which in both cases represents a significant reduction of the LSDA error relative to $\Delta$SCF predictions.

This sensitivity analysis provides a clear justification for the $\alpha$NKC$_0$ method that consists in setting the penalty coefficients $\alpha_{i\sigma}^{\rm NK}$ to be all equal to the same value, thereby avoiding in particular to treat negatively charged states. In explicit terms, $\alpha$NKC$_0$ consists in performing the following substitution in lieu of Eq.~\eqref{NK0}:
\begin{eqnarray}
&  v_{\rm Hxc}^{\rm KS}[\rho]   \nonumber \\
& \downarrow \label{ANK0} \\
& (1-\alpha)v^{\rm KS}_{\rm Hxc}[\rho] +\alpha v^{\rm KS}_{\rm Hxc}[\rho^{\rm NK}_{i\sigma}].  \nonumber
\end{eqnarray}

To complete the presentation of $\alpha$NKC$_0$, it is relevant to discuss the size consistency of the method. Within $\alpha$NKC$_0$, size consistency may be lost when considering a molecular system composed of separate fragments; in this situation, the $\alpha$ coefficient calculated for the global system may differ from the $\alpha$ coefficients that correspond to each of its subparts. (The same problem is known to arise in range-separated hybrid-DFT approaches where a single optimally tuned parameter $\gamma$ is employed to impose Koopmans' theorem.) An immediate solution to address this problem is to resort to NKC$_0$, which preserves the size consistency of the underlying local or semilocal functional owing to the orbital-specific coefficients $\alpha_{i\sigma}^{\rm NK}$. Note that other simple solutions may be employed, such as the approach that consists in using a different $\alpha$ for each separate molecular fragment. Following this procedure, orbitals localized on a given molecular fragment feel the same $\alpha$ while orbitals localized on different fragments have different $\alpha$'s, thereby preserving size consistency. 

\bibliography{article}

\end{document}